\documentclass{moriond}

\usepackage{graphicx}
\usepackage{hyperref}
\usepackage{amsmath, amssymb}
\usepackage{babel}
\usepackage{color}
\usepackage{mathrsfs}
\usepackage{slashed}

\def\be{\begin{equation}}
\def\ee{\end{equation}}
\def\bea{\begin{eqnarray}}
\def\eea{\end{eqnarray}}

\newcommand{\Photo}{\includegraphics[height=30mm]{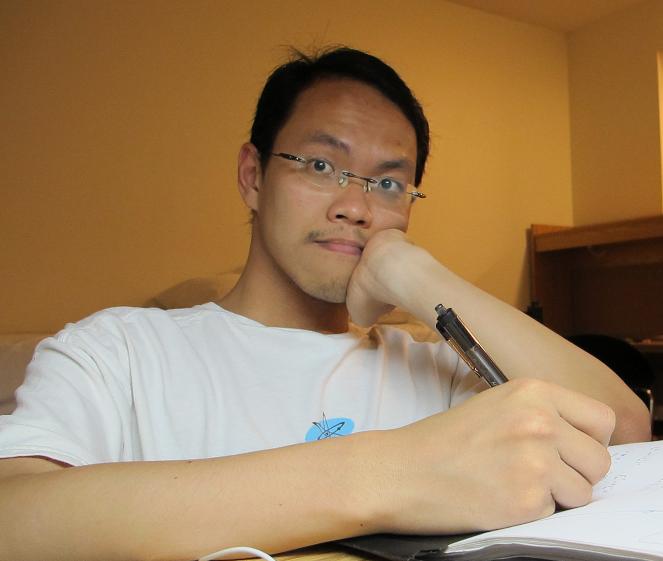}}

\begin{document}
\vspace*{4cm}
\title{$V_{ud}$ RADIATIVE CORRECTIONS WITH LATTICE INPUT}

\author{CHIEN-YEAH SENG}

\address{Helmholtz-Institut f\"{u}r Strahlen- und Kernphysik and Bethe Center for
	Theoretical Physics,\\ Universit\"{a}t Bonn, 53115 Bonn, Germany}

\maketitle\abstracts{
We will describe several pioneering efforts in the study of electromagnetic radiative corrections to semileptonic decay processes, with particular emphasis on the role of lattice QCD. These studies are essential for the precise extraction of the matrix element $V_{ud}$ from beta decays of pion, free neutron and $J^P=0^+$ nuclei, and are crucial to address several recently-emerged anomalies involving $V_{ud}$ and $V_{us}$, which may provide hints for physics beyond the Standard Model.}

\section{Anomalies in the top-row CKM matrix elements}

Several interesting anomalies that concern the
top-row elements in the Cabibbo-Kobayashi-Maskawa (CKM) matrix~\cite{Cabibbo:1963yz,Kobayashi:1973fv}
have recently emerged. First of all, since late 2018 one observes a 3$\sigma$ deviation from the unitarity relation required by the Standard Model (SM)~\cite{Zyla:2020zbs}:
\begin{equation}
|V_{ud}|^2+|V_{us}|^2+|V_{ub}|^2=0.9985(3)_{V_{ud}}(4)_{V_{us}}~,
\end{equation}
which is now frequently referred to in the literature as the Cabibbo angle anomaly (CAA). 
In the meantime, with the latest lattice calculation of the $K\pi$ form factor~\cite{Bazavov:2018kjg}, a previously-known anomaly in $V_{us}$ is further intensified:
\begin{equation}
|V_{us}|_{K_{\ell 3}}=0.2233(6)~,\:\:\:|V_{us}|_{K_{\mu 2}}=0.2252(5)~,
\end{equation}
where a 2.5$\sigma$ disagreement is observed between the values of $|V_{us}|$ extracted from the leptonic decay $(K_{\mu 2})$ and semileptonic decay $(K_{\ell 3})$ of the kaon. Although not yet conclusive, the anomalies above provide interesting hints for the existence of physics beyond the Standard Model (BSM).

An interesting feature of the aforementioned anomalies is that, their dominant sources of uncertainty come from the SM theory inputs rather than experiments. From an optimistic point of view, this implies that there is a high chance to discover BSM physics from these observables if one is able to significantly reduce all the major theory uncertainties. This is, however, extremely challenging because most of them are governed by Quantum Chromodynamics (QCD) in its non-perturbative regime, where analytic solutions do not exist. In this talk we will discuss several recent breakthroughs in the understanding of a particular class of SM corrections, namely the electroweak radiative corrections (RCs) that enter the beta decays of hadrons and nuclei and are crucial for the precise extraction of the top-row CKM matrix elements, especially $V_{ud}$. 

\section{Single-nucleon radiative correction}

\begin{figure}
	\begin{centering}
		\includegraphics[scale=0.6]{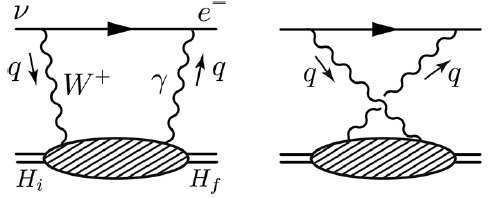}\hfill
		\par\end{centering}
	\caption{\label{fig:box}The $\gamma W$-box diagrams, where $H_i$ and $H_f$ are a pair of hadrons.}
\end{figure}

Having the largest magnitude, $V_{ud}$ plays the most important role in the top-row CKM unitarity. It is extracted (by far) most precisely from $0^+\rightarrow 0^+$ superallowed nuclear beta decays through the following master formula:
\begin{equation}
|V_{ud}|^2=\frac{2984.43~\mathrm{s}}{\mathcal{F}t(1+\Delta_R^V)}~,
\end{equation}
where $\mathcal{F}t$ is the $ft$-value corrected by nuclear-structure effects, while $\Delta_R^V$ denotes the single-nucleon RC. The latter has long been the dominant source of theory uncertainty, which originates from the single-nucleon axial $\gamma W$-box diagram (see Fig.\ref{fig:box}, with $H_i=n$ and $H_f=p$). These diagrams probe the hadron physics at all scales, in particular at the scale $q\sim 1$~GeV where effects of non-perturbative QCD dominates. 
A dispersion relation (DR) analysis of $\Delta_R^V$ in 2018: $\Delta_R^V=0.02467(22)$ (which gives $|V_{ud}|=0.97366(15)$)~\cite{Seng:2018yzq,Seng:2018qru} exhibits a significant shift of the central value from to the previous best determination in 2006: $\Delta_R^V=0.02361(38)$~\cite{Marciano:2005ec}, 
which was later confirmed by several independent studies~\cite{Czarnecki:2019mwq,Seng:2020wjq,Hayen:2020cxh,Shiells:2020fqp}. This shift is the main reason for the emergence of the CAA in the recent years, and therefore must be scrutinized with great care.

In a nutshell, the DR representation expresses the $\gamma W$-box diagram amplitude of a hadron $H$ as an integral of the function $M_H(Q^2)$, which is the first Nachtmann moment of the parity-odd, spin-independent structure function $F_3^{(0)}$ that involves the axial charged weak current and the isoscalar electromagnetic current:
\begin{equation}
\left.\Box_{\gamma W}^{VA}\right|_H=\frac{3\alpha_e}{2\pi}\int\frac{dQ^2}{Q^2}\frac{m_W^2}{m_W^2+Q^2}M_H(Q^2)~.
\end{equation} 
The function $M_H(Q^2)$ receives contributions from all on-shell hadronic states, some of which can be easily accounted for (such as the Born contribution), but most of them are not. In the case of nucleon (i.e. $H=N$), 
it turns out the most non-trivial contribution from multi-hadron intermediate states can be related to observables measured in inclusive neutrino-nucleus scattering processes through isospin symmetry~\cite{Seng:2018yzq,Seng:2018qru}, which provides the foundation for a data-driven analysis of $\left.\Box_{\gamma W}^{VA}\right|_N$. The limitation, however, comes from the low precision of the neutrino scattering data at $Q^2\sim 1$~GeV$^2$ which cannot efficiently constrain the physics it the non-perturbative regime~\cite{Bolognese:1982zd}. Better-quality data may come from the Deep Underground Neutrino Experiment (DUNE)~\cite{Acciarri:2016crz}, which is however not in reach in the near future. Therefore, first-principles calculations of $M_H(Q^2)$ with lattice QCD becomes an unavoidable task for further progress in this topic.

\section{First lattice QCD calculation}

\begin{figure}
	\begin{centering}
		\includegraphics[scale=0.6]{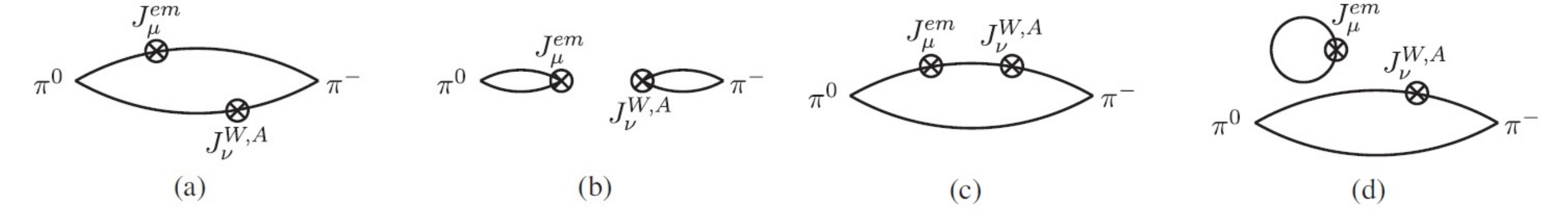}\hfill
		\par\end{centering}
	\caption{\label{fig:4pt}Quark contraction diagrams that correspond to $\mathcal{H}_{\mu\nu}^{VA}(x)$. Figures are taken from Ref.~\protect\cite{Feng:2020zdc}.}
\end{figure}

The first realistic lattice QCD calculation of hadronic $\gamma W$-box diagrams was performed by Xu Feng, Lu-Chang Jin and their group in early 2020 ~\cite{Feng:2020zdc}. They studied the simpler $\gamma W$-boxes of the charged pion as a prototype of the more complicated nucleon box diagrams that will be investigated in the future. Nevertheless, it is by itself interesting as $V_{ud}$ can also be measured in the semileptonic pion decay ($\pi_{e3}$) through the formula~\cite{Pocanic:2003pf}:
\begin{equation}
\Gamma_{\pi _{e3}}=\frac{G_F^2|V_{ud}|^2m_\pi^5|f_+^\pi(0)|^2}{64\pi^3}(1+\delta)I_\pi~,
\end{equation}
although its precision is severely limited by the large experimental uncertainties due to the small branching ratio. Among all the theory inputs to the equation above, the pion form factor
$f_+^\pi(0)$ and the kinematic factor $I_\pi$ are known to satisfactory precision, while the quantity $\delta$ which represents the electroweak RC was calculated with Chiral Perturbation Theory (ChPT)~\cite{Cirigliano:2002ng}:
$\delta=0.0334(10)_\mathrm{LEC}(3)_\mathrm{HO}$.
The main theory uncertainty comes from the poorly-constrained low energy constants (LECs) in ChPT, which practically originates from the $\gamma W$-box diagrams; a sub-dominant uncertainty comes from higher-order (HO) electroweak corrections.

To pin down the pion box diagrams, we need to know the function $M_\pi(Q^2)$ at all values of $Q^2$. At large $Q^2$ ($>$ 2 GeV$^2$) perturbative QCD works well, so we can adopt a leading-twist approximation with pQCD corrections:
\begin{equation}
	M_\pi(Q^2)=\frac{1}{12}\left[1-\sum_{i=1}^\infty c_n\left(\frac{\alpha_S}{\pi}\right)^n\right]
\end{equation}
where the coefficients $c_n$ are known up to $n=4$, which fully satisfy our precision requirement~\cite{Baikov:2010je}. On the other hand, at low $Q^2$, $M_\pi(Q^2)$ is obtained as a Euclidean spacetime integral:
\begin{equation}
M_\pi(Q^2)=-\frac{1}{6\sqrt{2}}\frac{\sqrt{Q^2}}{m_\pi}\int d^4x\omega(Q,x)\epsilon_{\mu\nu\alpha 0}x_\alpha \mathcal{H}_{\mu\nu}^{VA}(x)~,\label{eq:MpiQ2}
\end{equation}
where $\omega(Q,x)$ is a known function, and 
\begin{equation}
\mathcal{H}_{\mu\nu}^{VA}(x)=\left\langle \pi^0(P)\right|T[J_\mu^\mathrm{em}(x)J_\nu^{W,A}(0)]\left|\pi^-(P)\right\rangle
\end{equation}
is a four-point correlation function consists of quark contraction diagrams in Fig.\ref{fig:4pt} which were calculated directly on lattice. The calculation was done with five lattice QCD gauge ensembles (DSDR and Iwasaki gauge actions) at the physical pion mass generated by RBC and UKQCD Collaborations using 2+1 flavor domain wall fermions~\cite{Blum:2014tka}.

\begin{figure}
	\begin{centering}
		\includegraphics[scale=0.5]{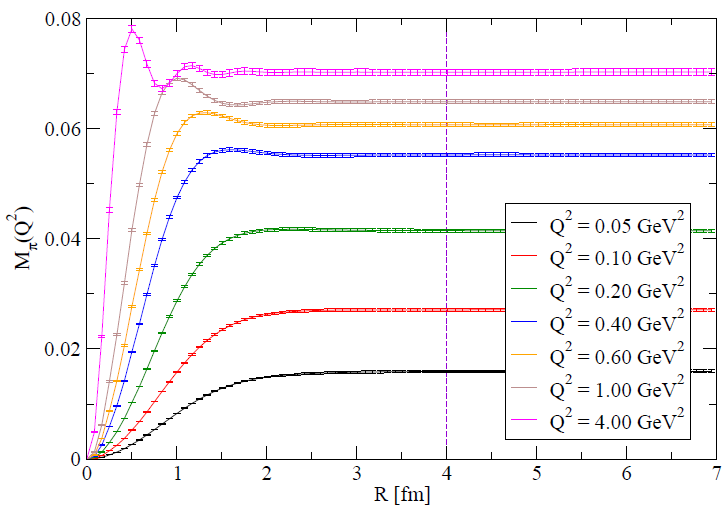}
		\includegraphics[scale=0.5]{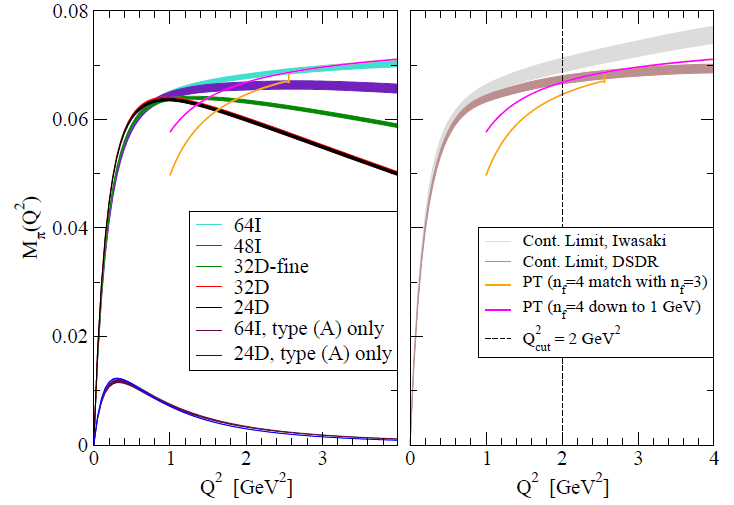}
		\hfill
		\par\end{centering}
	\caption{\label{fig:result}Main results of the first-principles calculation of $M_\pi(Q^2)$. Figures are taken from Ref.~\protect\cite{Feng:2020zdc}.}
\end{figure}

The main results are displayed in Fig.\ref{fig:result}. The left panel shows how fast the spacetime integral in Eq.\eqref{eq:MpiQ2} is saturated with the increase of the integration range $R$, and the right panel shows the lattice calculation of $M_\pi(Q^2)$ before and after the continuum extrapolation; the colored bands are lattice calculations while the lines are pQCD predictions. The full $\left.\Box_{\gamma W}^{VA}\right|_\pi$ is obtained by combining the pQCD result at $Q^2>2$~GeV$^2$ and the lattice result at $Q^2<2$~GeV$^2$. Effects of the main systematic uncertainties, such as the lattice discretization effect, the pQCD uncertainty and the higher-twist effects, are also properly taken into account.
The final result reads $\left.\Box_{\gamma W}^{VA}\right|_\pi=2.830(11)_\mathrm{stat}(26)_\mathrm{syst}\times 10^{-3}$,
where an overall 1\% precision is achieved.

The calculation above produces significant impacts to various aspects in beta decays. First, on the pion semileptonic decay, it reduces the theory uncertainty in the $\pi_{e3}$ RC by a factor of 3: $\delta= 0.0332(1)_{\gamma W}(3)_\mathrm{HO}$, and the updated value of $|V_{ud}|$ extracted from $\pi_{e3}$ now reads $
|V_{ud}|_{\pi_{e3}}=0.9740(28)_\mathrm{exp}(1)_\mathrm{th}$.
Despite still being plagued with large experimental uncertainties, this result is theoretically very clean. In fact, it now provides a major motivation for experimentalists to measure the $\pi_{e3}$ branching ratio with an order-of-magnitude increase in precision~\cite{ArevaloSnowmass}, which could eventually turn $\pi_{e3}$ into an avenue competitive to superallowed nuclear decays in the extraction of $V_{ud}$.

\begin{figure}
	\begin{centering}
		\includegraphics[scale=0.6]{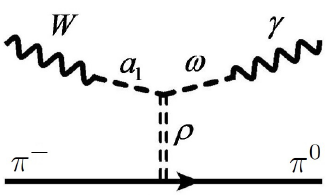}
		\includegraphics[scale=0.6]{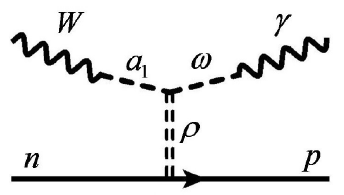}
		\hfill
		\par\end{centering}
	\caption{\label{fig:Regge}Multi-hadron contribution to $M_\pi(Q^2)$ (left) and $M_N(Q^2)$ (right) at low $Q^2$ described by a Regge-exchange picture. Figures are taken from Ref.~\protect\cite{Seng:2020wjq}.}
\end{figure}

Next, it turns out that the lattice calculation of the pion box diagrams also provide useful information to the single-nucleon box diagrams. The reason is that the multi-hadron contributions at low $Q^2$, which can be economically described by $t$-channel Regge exchanges (see Fig.\ref{fig:Regge}), is very similar between these two channels. One could then obtain one from the other upon replacing the coupling strength between the Regge trajectory and the target hadron. This provides an independent assessment of the single-nucleon RC: $\Delta_R^V=0.02477(24)$~\cite{Seng:2020wjq} which shows consistency with the DR result, but is based on a much more rigorous error analysis.

Finally, a very similar lattice calculation of the $K\pi$ box diagram in the flavor SU(3) limit was recently performed~\cite{Ma:2021azh}, which fixed some of the very important LECs in ChPT that enter the $K\rightarrow\pi$ semileptonic decay ($K_{\ell 3}$) which is one of the main channels to extract $|V_{us}|$. Combining this with a recently-proposed new treatment of the loop and bremsstrahlung corrections~\cite{Seng:2019lxf,Seng:2020jtz}, the theory uncertainty in the long-range electromagnetic RC to the $K\rightarrow\pi e^+\nu$ decay ($K_{e3}$) is reduced by an order of magnitude~\cite{Seng:2021boy,Seng:2021wcf} comparing to the previous state-of-the-art determination using ChPT~\cite{Cirigliano:2008wn}. This is physically significant in relation to a recently-proposed quantity $R_V=\Gamma_{K_{e3}}/\Gamma_{\pi_{e3}}$ that provides a new avenue to extract the ratio $V_{us}/V_{ud}$. With the reduction of the RC uncertainties in both $K_{e3}$ and $\pi_{e3}$, the quantity $R_V$ is theoretically very clean. Therefore, upon the future improvement of the $\pi_{e3}$ branching ratio measurement, $R_V$ will provide the most precise determination of $V_{us}/V_{ud}$ which will shed new lights on the $V_{us}$ anomaly. Notice that there are also proposals to calculate the full $K_{\ell 3}$ RC, including one-loop and bremsstrahlung contributions, directly on lattice~\cite{BoyleSnowmass}. These calculations, however, are more challenging and may take up to ten years to achieve a permille level precision comparable to the ChPT treatment.

\section{Future prospects}

An obvious future step is to calculate the nucleon $\gamma W$-box diagrams following exactly the same way as one does pion. There are several extra complications that make this task much more challenging: (1) The quark contraction becomes more complicated, (2) Much noisier data due to the exponentially-suppressed signal-to-noise ratio at large Euclidean time, and (3) The full control of systematic effects (e.g. excited-state contaminations) becomes more challenging. Nevertheless, there are already several groups, such as the Precision Neutron Decay Matrix Elements (PNDME) and Nucleon Matrix Elements (NME) Collaborations at Los Alamos, that have started initial investigations along this direction~\cite{BhattacharyaSnowmass}, so one may optimistically expect the first result to be available within the next one or two years.  

There is also an alternative proposal to approach the problem based on the application of the Feynman-Hellmann theorem on lattice~\cite{Seng:2019plg}, which has been successful previously in the study of P-even structure functions~\cite{Chambers:2017dov} but is not yet applied to beta decays. The underlying principle is rather straightforward: Upon introducing a periodic source term into the Hamiltonian,
\begin{equation}
H_\mathrm{src}(t)=2\lambda_1\int d^3x\cos(\vec{q}\cdot\vec{x})J_\mathrm{em}^2(\vec{x},t)-2\lambda_2\int d^3x\sin(\vec{q}\cdot\vec{x})J_A^3(\vec{x},t)~,
\end{equation}
the second-order perturbation to the nucleon energy is related to a convolution of the P-odd structure function $F_3(x,Q^2)$:
\begin{equation}
\left(\frac{\partial^2E_\lambda(\vec{p})}{\partial \lambda_1\partial\lambda_2}\right)_{\lambda=0}=\frac{4q_x}{Q^2}\int_0^1dx\frac{F_3(x,Q^2)}{1-\Omega^2x^2},\:\:\:\Omega=-2\vec{p}\cdot\vec{q}/\vec{q}^2~,\:\:\:\vec{q}^2=Q^2~,
\end{equation}
from which its first Nachtmann moment can be reconstructed through a linear combination with different choices of $\Omega$. This method may serve as a useful complement to the first approach, especially at large $Q^2$ where the direct calculation of four-point functions suffers from large lattice artifacts. In other words, it also provides a useful cross-check of the validity of the leading-twist + pQCD approximation of $M_N(Q^2)$ at $Q^2>2$~GeV$^2$. In short, it is always useful to tackle an important problem using several distinct methods in order to avoid unexpected systematic errors associated to specific approaches, and this of course applies to the study of box diagrams as well.   

\section{Conclusion}

The recently-emerged anomalies in the top-row CKM matrix elements call for a more careful look at the SM theory inputs, in particular the single-nucleon $\gamma W$-box diagrams that are responsible for the dominant theory uncertainty in the electroweak RCs to the free neutron and superallowed nuclear beta decays. In the lack of precise experimental data, one must rely on  first-principles calculations with lattice QCD to constrain the hadron physics in the non-perturbative regime that enters the box diagrams. 

The first-principles study of the pion $\gamma W$-box diagrams in early 2020 combining the results from 4-loop pQCD at high $Q^2$ and lattice QCD calculations at low $Q^2$ has achieved an impressive 1\% overall precision. 
Besides serving as a useful prototype for future calculations in the nucleon sector, it is physically interesting by itself as it removes the major theory uncertainty in $\pi_{e3}$ and prepares it as a future avenue for the high-precision extraction of $V_{ud}$. It also provides the first indirect lattice input for the nucleon box diagrams which confirms the correctness of the DR analysis. Furthermore, a similar calculation of the $K\pi$ box diagrams in the flavor SU(3) limit effectively constrains the poorly-known LECs in the ChPT and leads to a significant reduction of the RC uncertainties in $K_{e3}$, which may improve the extraction of $V_{us}$ and $V_{us}/V_{ud}$.  

Direct lattice calculations of the single-nucleon $\gamma W$-box diagrams are now on the way. They may have the last word on whether the existing CKM anomalies, in particular the CAA, are actually real.

\section*{Acknowledgments}

The author thank the organizers of the ``55$^\mathrm{th}$ Rencontres de Moriond'' for the kind invitation, and is grateful to X. Feng, M. Gorchtein, L. Jin, P. Ma and U.-G. Mei{\ss}ner for collaborations in various lattice-related projects.
The work of the author is supported in
part by the Deutsche Forschungsgemeinschaft (DFG, German Research
Foundation) and the NSFC through the funds provided to the Sino-German Collaborative Research Center TRR110 “Symmetries and the Emergence of Structure in QCD” (DFG Project-ID 196253076 - TRR 110, NSFC Grant No. 12070131001).

\end{document}